**An analytical model for the mechanical deformation of locally graphitized diamond**

M. Piccardo[1,2], F. Bosia[2, 3, 4(*)], P. Olivero[2,3,4], N. Pugno[5,6,7(*)]

[1]*Laboratoire de Physique de la Matière Condensée, CNRS-Ecole Polytechnique, 91128 Palaiseau Cedex, France*

[2]*Department of Physics and "Nanostructured Interfaces and Surfaces" inter-departmental centre, University of Torino, via P. Giuria 1, 10125 Torino, Italy*

[3]*Istituto Nazionale di Fisica Nucleare (INFN), Sezione di Torino, via P. Giuria 1, 10125 Torino, Italy*

[4]*Consorzio Nazionale Interuniversitario per le Scienze Fisiche della Materia (CNISM), via della Vasca Navale 84, 00146 Roma, Italy*

[5]*Laboratory of Bio-Inspired & Graphene Nanomechanics, Department of Civil, Environmental and Mechanical Engineering, Università di Trento, via Mesiano 77, 38123 Trento, Italy*

[6]*Center for Materials and Microsystems, Fondazione Bruno Kessler, via Sommarive 18, 38123 Povo, Italy*

[7] *School of Engineering and Materials Science, Queen Mary University of London, Mile End Road, London E1 4NS.*

(*) Corresponding authors:    e-mail: federico.bosia@unito.it , phone: +39 011 670 7889,

nicola.pugno@unitn.it , phone: +39 0461 282525

**Keywords**

**Abstract**

We propose an analytical model to describe the mechanical deformation of single-crystal diamond following the local sub-superficial graphitization obtained by laser beams or MeV ion microbeam implantation. In this case, a local mass-density variation is generated at specific depths within the irradiated micrometric regions, which in turn leads to swelling effects and the development of corresponding mechanical stresses. Our model describes the constrained expansion of the locally damaged material and correctly predicts the surface deformation, as verified by comparing analytical results with experimental profilometry data and Finite Element simulations. The model can be adopted to easily evaluate the stress and strain fields in locally graphitized diamond in the design of microfabrication processes involving the use of focused ion/laser beams, for example to predict the potential formation of cracks, or to evaluate the influence of stress on the properties of opto-mechanical devices.

## 1. Introduction

A relevant number of works has concentrated in recent years on the application of MeV-ion-induced graphitization to fabricate and functionalize microstructures and devices in single-crystal diamond, including bio-sensors [1], ionizing radiation detectors [2, 3], bolometers [4], nano-electromechanical systems (NEMS) [5, 6], photonic structures [7-10] and optical waveguides [11, 12]. Laser-induced graphitization has also been employed to fabricate metallo-dielectric structures [13] and ionizing radiation detectors [14] in diamond. This versatility is due to the fact that both MeV-ion and laser focused beams can locally deliver high power densities in specific regions within the diamond bulk with micrometric spatial resolution in all directions, thus creating confined regions where the diamond lattice structure is critically damaged. In these regions, annealing leads to the graphitization of the damaged structure, whilst the remaining surrounding material is largely restored to pristine diamond, so that well-defined structures can be created by selectively etching the graphitized regions [3, 5-12] or taking advantage of the optical/electrical properties of the graphitized regions [1, 2, 4, 13, 14]. At significantly lower damage densities (i.e. well below the graphitization threshold), ion implantation was employed to tailor the optical properties of diamond either by modifying its refractive index [15-18] to directly write/fine-tune waveguiding structures [19] and photonic structures [20], or to induce spectral shifts in the emission of luminescent centres [21]. In all of these cases, accurate knowledge is required of the modification of the diamond lattice structure as a function of implantation/irradiation parameters and *in-situ*/post-processing annealing conditions, in order to exactly localize the graphitized/modified layer and predict its structural effects on the surrounding material.

As far as ion implantation is concerned, the critical damage level ($D_C$) above which diamond is subject to permanent amorphization and subsequent graphitization upon thermal annealing is referred to as the "graphitization threshold" [22], and its dependence on implantation parameters has been ascertained (e.g. depth and/or local strain, self-annealing, etc.) [23-27]. An observable effect of ion implantation and laser irradiation in diamond is surface swelling, due to the density variation in the sub-superficial damaged regions and the corresponding constrained volume expansion [28-30]. It is therefore possible to analyze this effect in order to infer the structural modifications occurring in ion-implanted diamond and the extent of the density variation. In previous studies a phenomenological model accounting for saturation in vacancy density was developed, and finite-element (FEM) simulations were performed to compare numerical results with experimental surface swelling measurements [31-33]. The use of FEM modelling requires the use of specialized software and specific expertise in the field. On the other hand, oversimplified mechanical models have often been used to calculate mechanical deformations[28] and strains[34] in ion-implanted diamond in the literature, with limited predictive capabilities. In this paper, we propose a more rigorous analytical approach to derive material swelling and internal stresses following the laser or MeV-ion irradiation of diamond, and validate it by comparing its predictions to experimental and numerical data in a number of studies.

## 2. Analytical model

### 2.1    2D modelling of graphitic layers within the diamond crystal

Well-defined graphitic regions can be created in diamond either by MeV ion implantation followed by high-temperature annealing or by irradiation with high-power pulsed laser beams. As shown in Fig. 1, the irradiation of a crystalline structure with light MeV ions at suitable fluences generally results in the formation of a sub-superficial amorphized layer, due to the peculiar depth profile of the ionic nuclear energy loss. For a given material, the thickness and depth of the amorphized layer primarily depends on the ion species and energy, as well as implantation fluence. It is worth noting that the volumetric vacancy density reported in Fig. 1 was estimated by assuming a simple linear dependence from the implantation fluence, i.e. by multiplying the fluence by the linear vacancy density per ion $\lambda$, as generated by SRIM-2008.04 Monte Carlo simulations [35] in "Detailed calculation" mode and by setting an atomic displacement value of 50 eV [36]. It has been shown that such a crude approximation (that neglects non-linear damage effects such as defect-defect interaction and self-annealing) does not provide a physically plausible estimation of the vacancy concentration [28], nonetheless it is suitable to describe the depth and thickness of graphitized layers in samples after high-temperature annealing, provided that the correct empirical graphitization threshold is adopted [34]. Moreover, as mentioned in the previous section, the diamond layer that has been damaged above the graphitization threshold is assumed to be converted to graphite upon high-temperature (i.e. 1200 °C) annealing, whilst the remaining upper layer is assumed to have reverted to the pristine diamond phase. The latter assumption is only partially justified, since it has been established that even after high-temperature annealing, the crystal structure of implanted diamond retains a small degree of residual damage [27], as clearly observable in the electrical characterization of the material [2]. However, this effect can reasonably be neglected when considering variations in the mechanical properties[26, 37].

Although this second strategy is not considered in the present work, it is worth mentioning that extended sub-superficial graphitic layers in diamond can be obtained upon high-power pulsed laser irradiation [38, 39] through non-equilibrium photo-induced phase transitions induced by fast electronic excitations that change their chemical potential [40]. In this case, a post-irradiation annealing step generally is not necessary to finalize the conversion to a graphitic phase, and the depth and thickness of the buried graphitic layer is directly determined by the extension of the region scanned by the focal point of the laser beam within the sample.

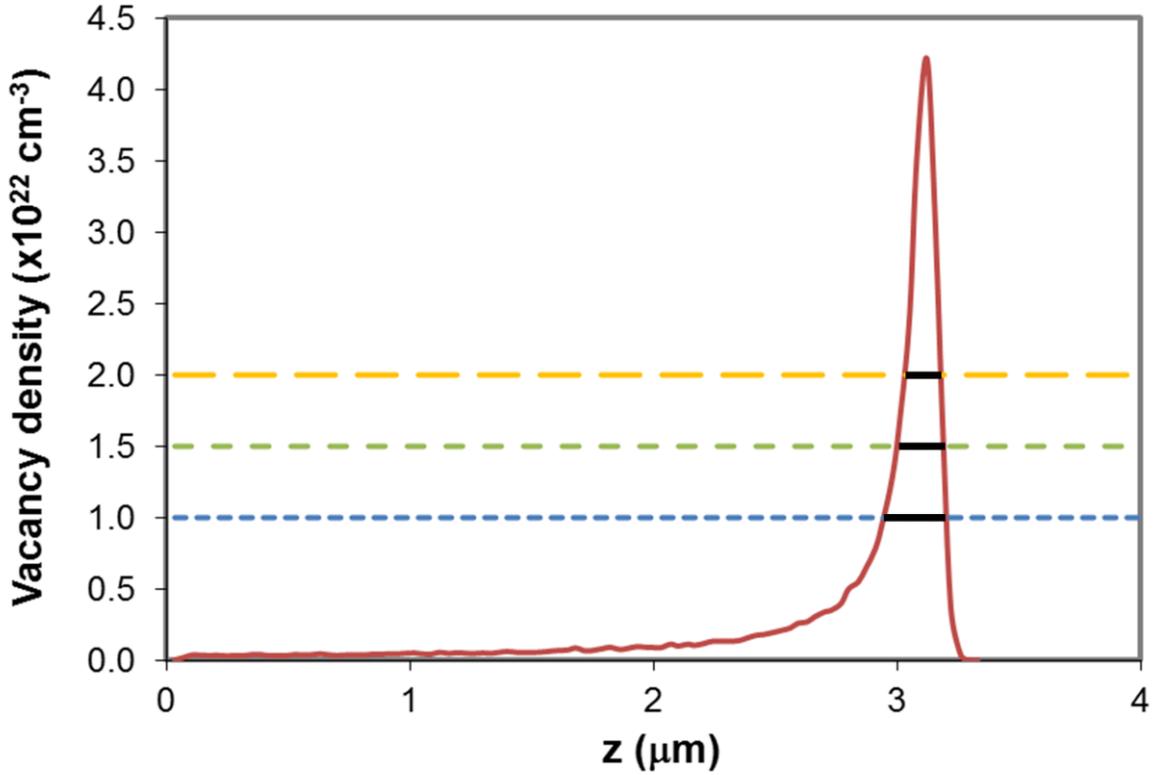

*Figure 1: Depth profile of the vacancy density induced in diamond from 1.8 MeV He⁺ ions implanted at a fluence of $2 \times 10^{16}$ cm⁻², as derived from the application of a linear fluence dependence to the numerical output of SRIM simulations. As an example, three critical damage thresholds are plotted (dashed lines) leading to different estimations of the thickness value h of the graphitized layer ( black horizontal segments within the damage profile peak).*

Regardless of the graphitization strategy, let us consider a diamond sample with a rectangular ion- or laser-irradiated area of length *l* and width *w*. The cross-sectional geometry of the sample is shown in Fig. 2a. The sample is modelled as a two-layer structure: a pristine diamond beam resting (with thickness *t*) on a graphitic elastic foundation (of thickness *h*), the latter undergoing a constrained expansion, due to its decrease in mass density from diamond to graphite. As a first approximation, the arbitrarily extended diamond crystal surrounding the lateral sides of the graphitic region (the "insert") is assumed to be infinitely rigid. This prevents lateral expansions, so that displacements are purely vertical (i.e. in the *z* direction, Fig.2b). In order to perform an analytical study of the deformation of the diamond surface layer due to expansion of the underlying graphitic region, we employ the equation of a beam on a Winkler foundation, deriving it from the elastic beam equation [41], where the diamond and graphitic layers respectively correspond to the two above-mentioned components.

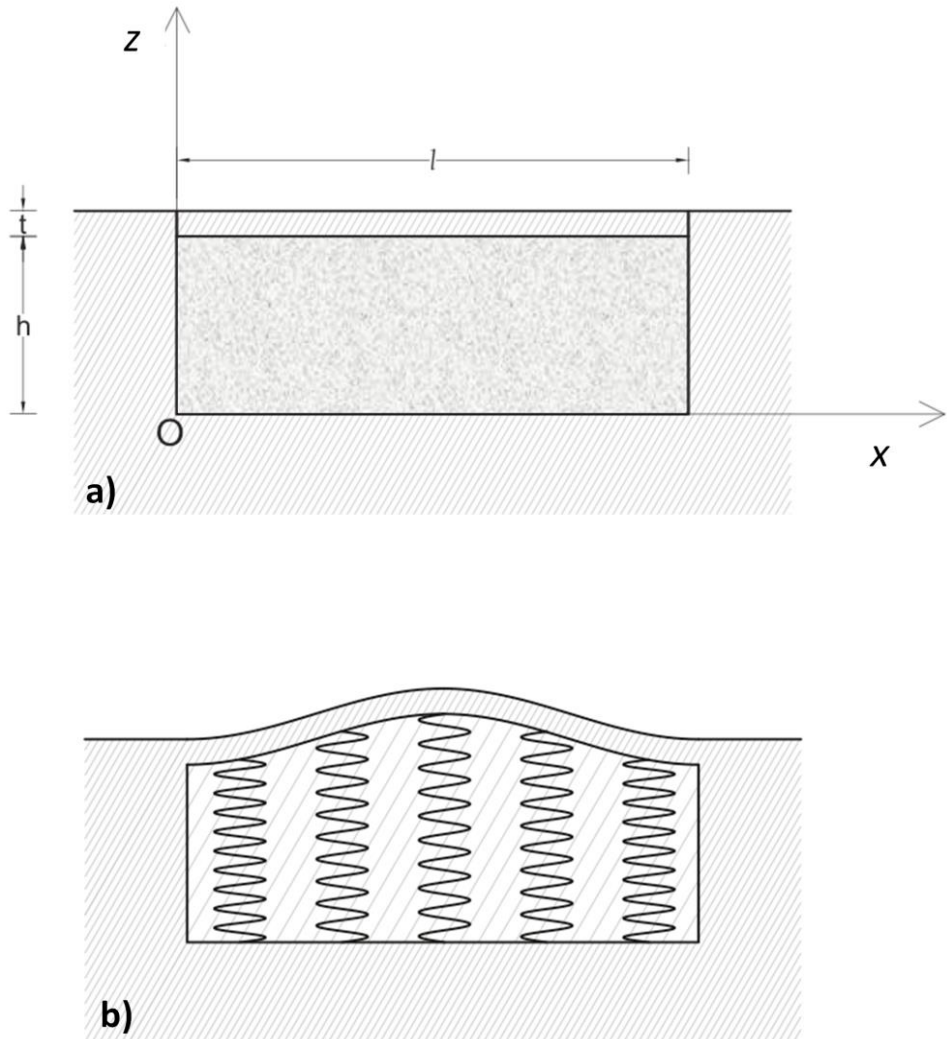

*Figure 2: a) Schematic representation of a two-dimensional section of the locally graphitized diamond region; b) Deformed shape of the implanted region, modelled as an elastic foundation. The images are not to scale.*

The superficial swelling of the diamond beam is thus due to the expansion of the graphitic elastic foundation because of its decrease in density. The two regions are assigned different Young's moduli (i.e. $E_d$ for diamond and $E_g$ for graphite), and the density decreases from the initial diamond value $\rho_d$ to the graphite one $\rho_g$.

The deformation of the top diamond layer is thus calculated using the Euler–Bernoulli elastic beam equation [41]:

$$\frac{\partial^4 v}{\partial x^4} = \frac{q(v)}{E_d \cdot I} \tag{1}$$

where $x$ is the horizontal coordinate (see Fig. 2a), $v$ is the layer deformation in the $z$ direction, $I = w \cdot t^3 / 12$ is the moment of inertia of the layer, and $q$ is the load per unit length applied to the layer.

Here, we adopt the Euler–Bernoulli formulation instead of Timoshenko beam theory, since $E_d I / \left( K L^2 t w G_d \right) \ll 1$, where $K$=5/6 and $G_d$ is the diamond shear modulus, which amounts to supposing that shear effects are negligible[42]. Thus, the graphitic layer (elastic foundation) exerts a load per unit length $q$ along the diamond layer, with:

$$q(v) = -k_g \cdot \left( v - v_0 \right) \tag{2}$$

where $v_0$ is the unconstrained elongation of the graphitic foundation in the $z$ direction, and $k_g$ is its stiffness, which is in turn given by:

$$k_g = \frac{E_g \cdot w}{h + v_0}$$ (3)

If we neglect the total mass of the implanted ions (whose contribution can be estimated to correspond at most to 1% of the total mass of the region under consideration), the volume variation in the implanted layer is primarily determined by its density variation. By considering a purely vertical expansion (due to the constraining effect of the surrounding pristine diamond region), we obtain, for finite variations:

$$\frac{\Delta\rho}{\rho} = -\frac{\Delta V}{V} = -\frac{\Delta h}{h}$$ (4)

Therefore, we have:

$$v_0 = \Delta h = -\frac{h \cdot \Delta\rho}{\rho_d} = \frac{h \cdot (\rho_d - \rho_g)}{\rho_d}$$ (5)

Using Eqs. (2) and (5), Eq. (1) becomes:

$$\frac{\partial^4 v}{\partial x^4} + \kappa \cdot v = \kappa \cdot v_0 \qquad (6)$$

with

$$\kappa = \frac{E_g \cdot 12}{E_d \cdot (h + v_0) \cdot t^3} ; \qquad (7)$$

This 4$^{\text{th}}$-order differential equation can be solved for given $\kappa$ and $v_0$ values, using the theory of beams on elastic supports [41], thus obtaining:

$$v(x) = e^{\beta \cdot x}[c_1 \cdot \cos(\beta \cdot x) + c_2 \cdot \sin(\beta \cdot x)] + e^{-\beta \cdot x}[c_3 \cdot \cos(\beta \cdot x) + c_4 \cdot \sin(\beta \cdot x)] + v_0 \qquad (8)$$

where $\beta = \sqrt[4]{\dfrac{\kappa}{4}}$ and the coefficients $c_1$, $c_2$, $c_3$, $c_4$ are calculated by applying the following "clamped-clamped" boundary conditions:

$$v(0) = v(l) = \frac{\partial v}{\partial x}(0) = \frac{\partial v}{\partial x}(l) = 0 \qquad (9)$$

which imply that i) no deformations occur at the borders of the implanted layer and ii) the derivative of the deflection function is zero at the above-mentioned points. The principal stress in the $z$ direction in the implanted area $\sigma_z(x)$ can be calculated from Eqs. (2) and (3), as:

$$\sigma_z(x) \approx \frac{q(x)}{w} = \frac{E_g \left[ v_0 - v(x) \right]}{h + v_0} \tag{10}$$

while the stress components in the perpendicular directions are:

$$\sigma_y(x) = \sigma_x(x) = \frac{\upsilon}{1-\upsilon} \frac{q(x)}{w} = \frac{\upsilon}{1-\upsilon} \frac{E_g \left[ v_0 - v(x) \right]}{h + v_0} \tag{11}$$

where $\upsilon$ is the Poisson's ratio of diamond.

## 2.2    2D modelling of non-uniformly damaged layers in diamond

As-implanted (i.e. not subsequently annealed) samples after MeV ion irradiation represent the ideal system to test a scenario in which the diamond structure is subjected to a non-uniform damage profile, with regions at different depths being characterized by graded damage densities. Due to the characteristic nuclear energy loss profile of MeV ions in matter (see Fig. 1), the implanted layer will display a non-uniform depth profile $\lambda(z)$ of the linear vacancy density $z$, with a typical end-of-range peak. As mentioned above, the volumetric vacancy density depth profile $\rho_V$ can be obtained from the SRIM code output [35] in a linear approximation, i.e. by assuming $\rho_V = F \cdot \lambda(z)$. It is worth noting that in the pre-annealing case corrections are needed for high-fluence implantations, in order to account for damage saturation effects. These are discussed extensively in previous works [31-33]. Following the above-mentioned results, we assume that the mass density profile can be calculated as:

$$\rho(F, z) = \rho_d - (\rho_d - \rho_{aC})(1 - e^{-\frac{\lambda(z)F}{\alpha}})$$

(12)

where $\rho_{aC}$ is the amorphous carbon density, $F$ is the implantation fluence and $\alpha$ is an empirical parameter depending on the implantation conditions that accounts for the defect recombination probability [32, 33]. The mass density variation profile corresponding to a 1.8 MeV He$^+$ implantation at a fluence of $1 \times 10^{16}$ cm$^{-2}$ was calculated by assuming $\rho_{aC} = 2.14$ g cm$^{-3}$ [34] and $\alpha = = 7 \times 10^{22}$ cm$^{-3}$ [32], as shown in Fig. 3.

The analytical model described in the previous section can be extended to account for this scenario as well. To apply the formalism introduced in Section 2.1, we assume as a first approximation that surface deformations due to a density profile such as that in Fig. 2 are equivalent to those induced by a buried layer with uniform average density $\overline{\rho}$. For practical purposes, the buried layer is defined as a region where a >1% mass density variation is induced with respect to the undamaged crystal. The "cap" region between the surface and the buried damaged layer plays the role of the "beam" in this case. Then, substituting the average density $\overline{\rho}$ in place of $\rho_g$ in Eq. (5), the formalism presented in Section 2.1 can be extended to model the case of an as-implanted layer.

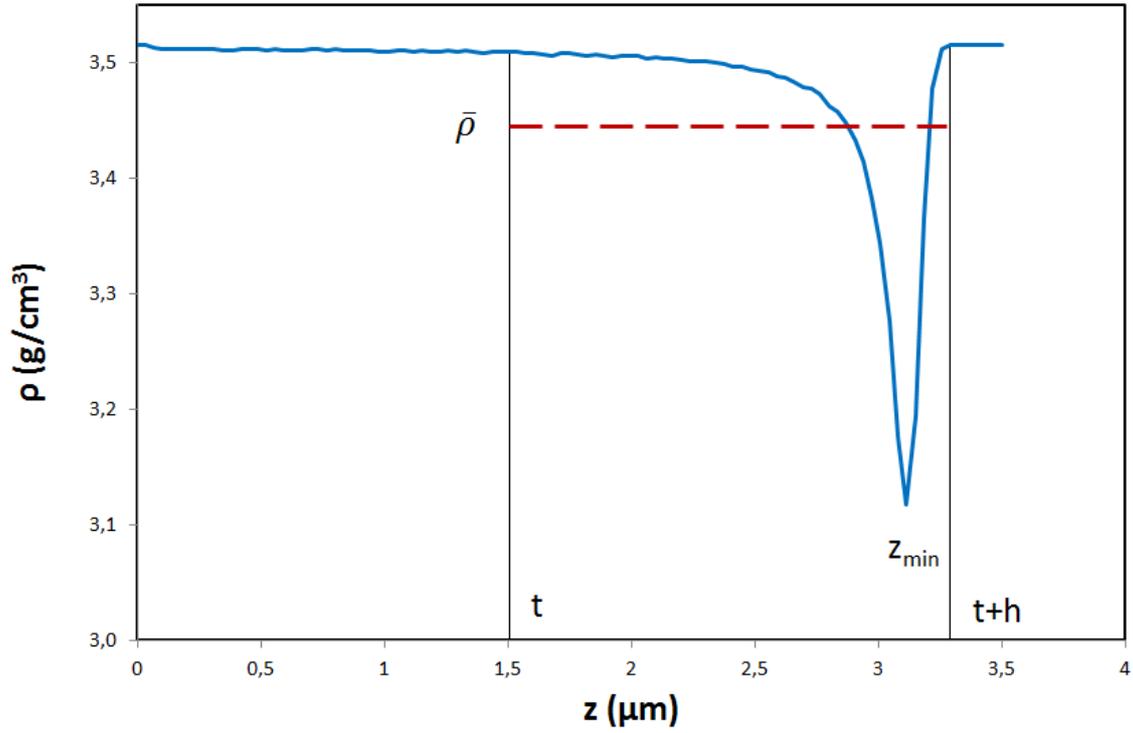

*Figure 3: Calculated depth profile of the mass density for a 1.8 MeV He⁺ implantation at a fluence of 1×10¹⁶ cm⁻², and corresponding average density in the implanted region ( $\bar{\rho}$ = 3.45 g cm⁻³).*

### 2.3    3D modelling of implanted layer

The model derived in Sections 2.1 and 2.2 can also be extended to a three-dimensional geometry, considering the equation of plates on elastic foundation [41]:

$$\frac{\partial^4 w}{\partial x^4} + \frac{\partial^4 w}{\partial y^4} + 2\frac{\partial^4 w}{\partial x^2 \partial y^2} = \frac{q}{G}$$

(13)

where $x$ and $y$ are the coordinates across the plane defined by the sample surface, $w$ is the vertical plate deformation, $q = -k(w - w_0)$ is the load along the $z$ direction and $G$ is the flexural rigidity of the plate. For simplicity, we assume independent $x$ and $y$ deformations and look for a solution in the form:

$$w(x, y) = \sqrt{v(x) \cdot v(y)}$$

(14)

where $v$ is the two-dimensional "beam" deformation. Consequently, we can derive an analytical three-dimensional expression for the swelling profile. Due to the simplified nature of Eq. (14), the expression is probably unsuitable for calculating edge effects, however it is acceptable as a first approximation, as verified in the following Sections.

## 3. Results

### 3.1    *MeV ion implanted and thermally annealed samples*

In order to validate the model outlined in the previous Section, we compare analytical predictions with experimental data and Finite Element Model (FEM) numerical simulations available from previous works for He$^+$ implantations at various fluences [32, 33]. Ion implantation was performed on HPHT (produced by Sumitomo, type Ib, (1 0 0) crystal orientation) samples, 3×3×0.5 mm$^3$ in size, with two optically polished opposite large faces. The samples were irradiated with 1.8 MeV He$^+$ ions at the ion microbeam line of the INFN Legnaro National Laboratories. Typically, ~125×125 μm$^2$ square areas were implanted by raster scanning an ion beam with size of 20-30 μm. The implantation fluences, ranging from 1×10$^{16}$ cm$^{-2}$ to 2×10$^{17}$ cm$^{-2}$, were controlled in real time by monitoring the X-ray yield from a thin metal layer evaporated on the sample surface. The implantations were performed at room temperature, with ion currents of ~1 nA. Surface swelling data were acquired at the Istituto Nazionale di Ottica (INO) with a Zygo NewView 6000 system, which exploits white light interferometry to provide detailed, non-contact measurements of 3-D profiles [43]. FEM simulations were carried out using the "Structural mechanics" module in COMSOL Multiphysics ver. 4.3 [44]: basically, the local density variation is modelled as a constrained volume expansion, similar to a thermal expansion problem, as reported in [31, 32]. Material properties for calculations were taken from literature, as done in previous works [21, 33, 45]: $E_d$ =1220 GPa, $E_g$ =14 GPa, $\upsilon$= 0.2, $\rho_d$ = 3.5 g cm$^{-3}$, $\rho_g$ = 2.1 g cm$^{-3}$. For simplicity, elastic properties are supposed homogeneous and identical in all crystal directions. A decisive parameter in this case is the graphitization threshold, $D_C$, which has been found to vary significantly as a function of implanted ion species and energy, implantation fluence and temperature, etc.[23-27, 34, 46-51]. Different $D_C$ values imply different thicknesses of the graphitized layer. Figure 1 illustrates how an increase of the threshold $D_C$ corresponds to a reduction in thickness $h$.

Analytical calculations of surface deformation (taken at the centre of the implanted area) are therefore carried out for various $D_C$ values, so as to evaluate the dependence from this parameter and identify the value which has closest adherence to the experimental results. Results are shown in Fig. 4, with $D_C$ varying between $1.0 \times 10^{22}$ cm$^{-3}$ and $2.0 \times 10^{22}$ cm$^{-3}$. From the comparison with experimental results, the closest adherence to experimental data is obtained for $D_C$ = $1.5 \times 10^{22}$ cm$^{-3}$, which is compatible with values obtained for the same type of implantations in previous works [32]. It is worth noting that both experimental and calculated values display a threshold fluence value below which there is no surface swelling, which corresponds to the fluence value for which the peak in the vacancy density curve remains below $D_C$, and thus thermal annealing induces the reconversion of all damaged carbon to diamond. Taking this value for $D_C$, we now compare the analytically calculated values to those obtained with FEM numerical simulations carried out using the same parameters, in order to evaluate the reliability of the model introduced in Section 2. Results are also shown in Fig. 4. Analytical values tend to systematically slightly overestimate FEM values. This is attributed to the above-mentioned approximation of "rigid" (non-deforming) lateral sides, however discrepancies are small (< 11%), thus proving the validity of the analytical approach and of the relevant simplifying hypotheses.

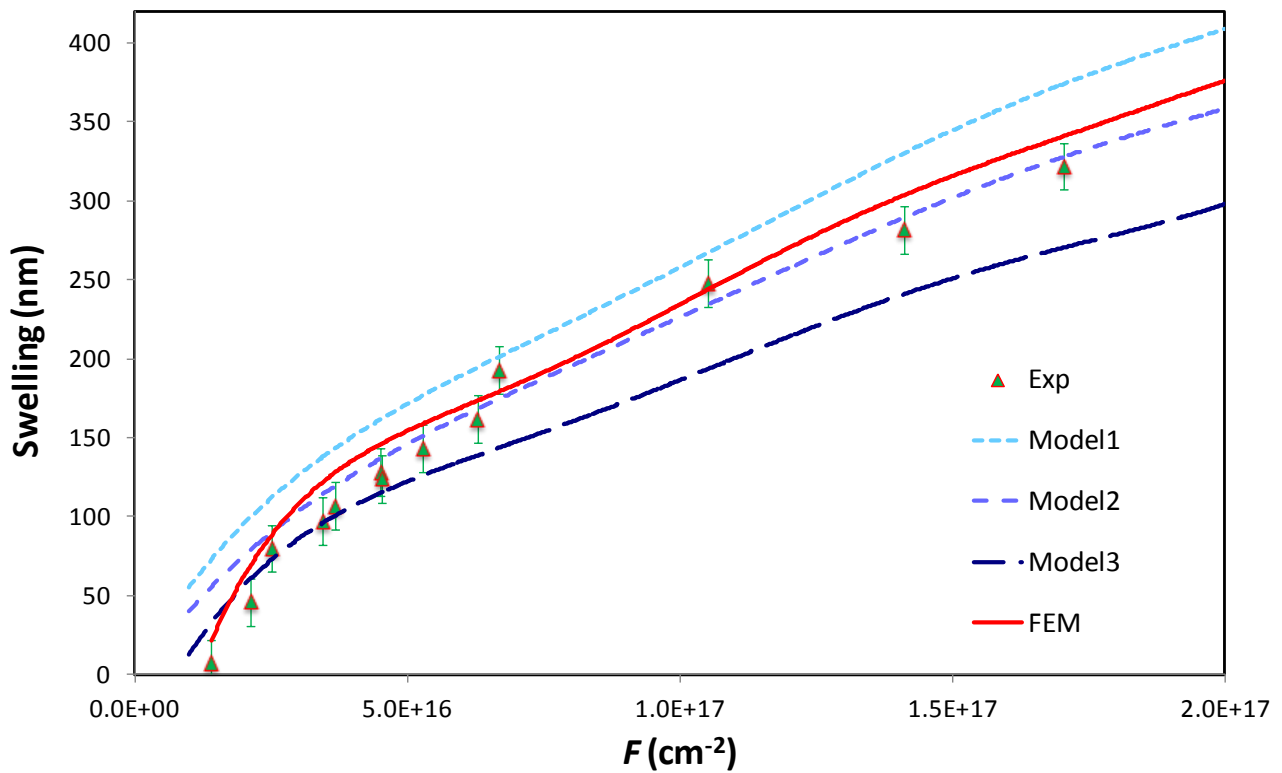

*Figure 4: Experimentally measured ("Exp."), simulated ("FEM"") and analytical ( "Model") surface swelling vs fluence for 1.8 MeV He⁺ implantations after 1200 °C annealing. The three analytical curves are shown for different values of the critical damage threshold: Model1: $D_C$ =1.0×10²² cm⁻³, Model2: $D_C$ =1.5×10²² cm⁻³, Model3: $D_C$ =2.0×10²² cm⁻³. FEM results are only presented for $D_C$ = 1.5×10²² cm⁻³.*

The reliability of the model can also be assessed by comparing experimental data with analytically and numerically calculated values for the surface swelling profile over the whole width of the implanted area for a given fluence value, in this case $F$=3×10$^{16}$ cm$^{-2}$ (Fig. 5a). Again, discrepancies are small (< 6%), but edge effects are slightly different in the three cases, due to the approximation of rigid lateral sides and the effect of the Gaussian profile of the employed ion beam.

Using Eqs. (10) and (11), stresses along the implanted area can also be analytically calculated. As reported in Fig. 5b, where the lateral distribution of analytically calculated stresses at the surface is shown, stresses are particularly pronounced at the edges of the implanted layer. These are the locations of most probable fracture initiation for high implantation fluences, particularly when the implantation depth is small. In this respect, the present calculation procedure can provide a simple and rapid tool to verify when a chosen graphitization process presents the risk of cracking at the graphite-diamond interface. For this purpose, it is possible to adopt one of the several established mechanical failure criteria, e.g. the Von Mises yield criterion [52], which can be expressed as a function of principal stresses:

$$\left(\sigma_x - \sigma_y\right)^2 + \left(\sigma_y - \sigma_z\right)^2 + \left(\sigma_z - \sigma_x\right)^2 \geq 2\sigma_Y^2 \qquad (15)$$

Where $\sigma_Y$ is the yield stress for the material, which in the case of diamond is approximately 130 GPa [53].

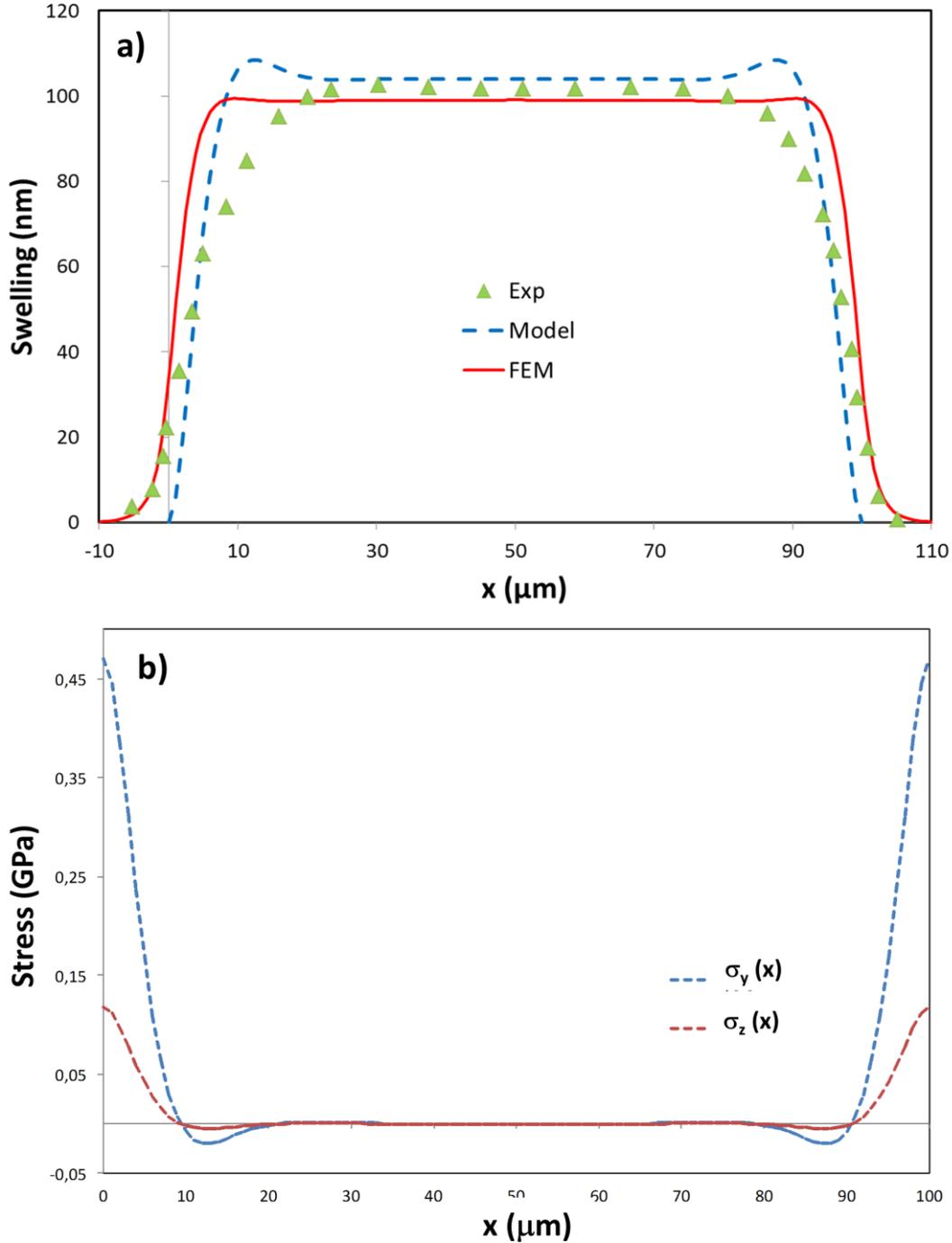

*Figure 5: a) Experimental ("Exp") and analytically ("Model") and numerically ("FEM") calculated surface deformation profiles for 1.8 MeV He$^+$ implantations (F = 3×10$^{16}$ cm$^{-2}$) after full annealing, as a function of the lateral coordinate. Due to the assumption of infinitely rigid lateral confinement the analytical prediction gives a higher swelling peak as compared to the FEM simulation. b) Corresponding principal stresses in the surface layer, calculated from analytical values.*

*3.2   As-implanted samples*

In the case of the as-implanted diamond substrate, experimental and numerical FEM data are also available [31, 33] and can be used to check the validity of the proposed approach. In this case, the free parameters of the model are $\alpha$ (i.e. the parameter describing the defect recombination probability [31]), and $t$ (i.e. the approximate thickness of the "cap" layer in which the effect of ion-induced damage is negligible). Another relevant parameter in the calculation is the limiting density of the ion-damaged material, i.e. $\rho_{dC}$ =2.14 g cm$^{-3}$, as determined in [34]. Again, a parametric study was carried out to determine the values of the parameters $t$ and $\alpha$ yielding the best adherence to experimental data. The values $t$ = 1.5 µm and $\alpha$ = 7×10$^{22}$ cm$^{-3}$ were obtained, in good agreement with previous studies [32]. Results are shown in Fig. 6a, where the surface swelling at the center of the implanted area is reported as a function of the implantation fluence. As for annealed samples, analytical results are compared to experimental data and FEM numerical results. As previously, analytical values tend to slightly overestimate FEM values, but the agreement is satisfactory, and discrepancies with experimental data remain below 10% for medium/low fluence values. At high fluences, the agreement between experimental and analytical/numerical datasets is worse than in the case of annealed samples. This is attributed to the additional approximation of using an average equivalent density for the implanted layer (see Section 2.2). Also, as for annealed samples, the analytical expression of the swelling profile along the lateral direction is used in Fig. 6b for a comparison with FEM simulations for a $F$=3×10$^{16}$ cm$^{-2}$ fluence. Again, analytical predictions slightly overestimate the FEM values, consistently with the assumption of infinitely stiff lateral confinement, leading to a more pronounced vertical deformation of the surface.

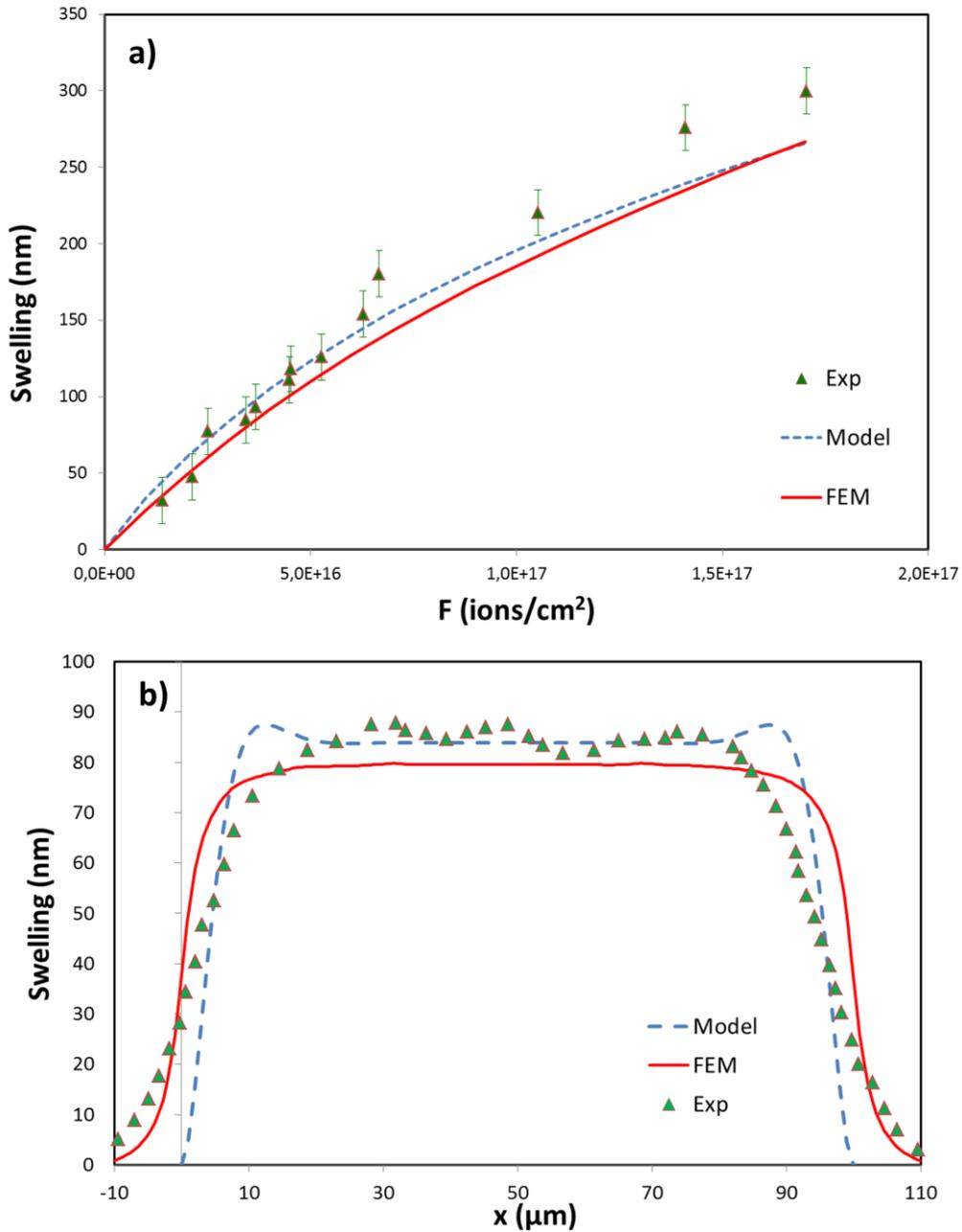

*Figure 6:a) Experimental, analytical and numerical swelling values for 1.8 MeV He$^+$ implantations in as-implanted samples as a function of implantation fluence. b) Corresponding deformation profiles for a fluence of F=3×10$^{16}$ cm$^{-2}$.*

## 3.3 3-D deformation profiles

Figure 7 reports a comparison between the results of three-dimensional analytical modelling and experimental measurement of the surface deformation resulting from a 1.8 MeV He$^+$ implanted $250 \times 250 \mu m^2$ area at a fluence $F = 2 \times 10^{16}$ cm$^{-2}$ and after thermal annealing. The overall agreement is good, although edge effects are clearly not adequately accounted for in the analytical model. The decrease to zero swelling values at the edges of the implanted area is found experimentally to be much more gradual than predicted using the model, as for 2D modelling results. This could also be due to the Gaussian-like decay in intensity of the ion microbeam at the edges of the implanted region, and possibly to ion-straggling effects through the depth. Overall, we can conclude that analytical predictions are generally reliable away from the edges of the implanted regions. In the proximity of the edges, more sophisticated modeling would be required to correctly draw definite conclusions, however the values predicted with the present model can provide an upper bound for stress calculations and failure predictions in this region.

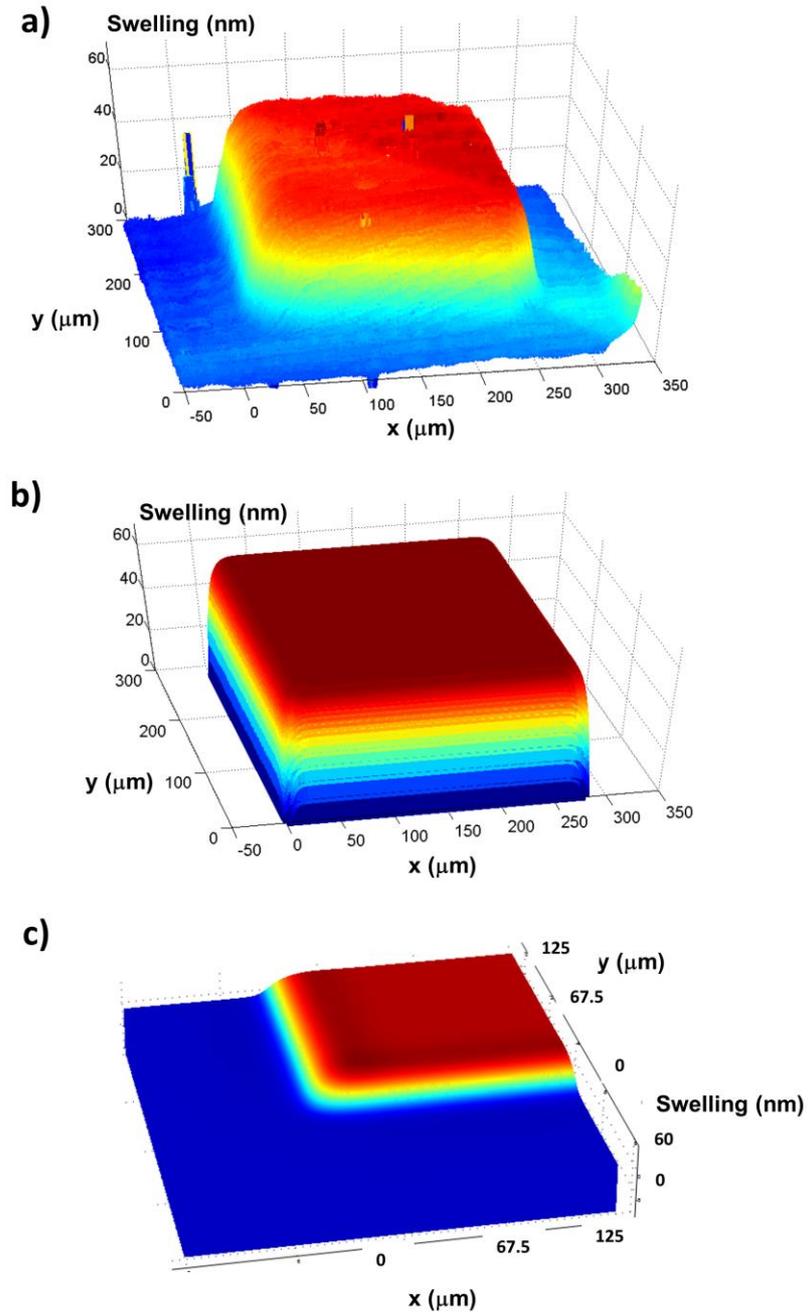

*Figure 7: a) Experimentally measured, b) analytically-calculated and c) FEM calculated 3-D swelling profiles relative to a 250×250 μm² area implanted with 1.8 MeV He⁺ ions at fluence F = 2×10¹⁶ cm⁻² after 1200 °C thermal annealing. Colour scale is the same for the three images. For the FEM profile, due to symmetry, only one quarter of the sample is shown.*

## 4. Conclusions

An analytical model was developed to predict the surface deformation and internal stresses in single-crystal diamond samples which have undergone MeV ion implantation. The predictions of the two-dimensional analytical model have been compared with experimental and numerical FEM data available from literature for 1.8 MeV He$^+$ implantations at various fluences, both for as-implanted and 1200 C annealed samples. The free parameters of the model have been optimized in order to maximize the adherence with experimental data, and have been found to be consistent with previous studies. Analytical results generally display good agreement with experiments and FEM simulations, in particular for the post-annealing case where the mass density profile is constituted of two distinct and homogeneous layers, i.e. a diamond "beam" resting on a graphitized "foundation". The systematic slight overestimation of the model with respect to the numerical calculations could be corrected by eliminating the assumption of infinitely stiff lateral confinement, although this would significantly complicate the mathematical derivation. The model was also extended to three dimensions using plate theory, allowing a direct comparison of the analytical swelling surface with experimental data. In future, the model can be extended to account for intermediate thermal treatments below the temperature of full graphitization, so as to allow the comparison with a wider range of experimental data. Since best fitting parameters have also been derived in the present work for He MeV implantations, the model can now be used as a predictive tool in the design of ion-beam/laser microfabrication procedures in diamond based on damage-induced graphitization, particularly by predicting undesired mechanical effects such as cracks, as well as other mechanical effects on the optical properties of the material (refractive index, spectral shifts in the emission from luminescent centers, birefringence, etc.).

**Acknowledgments**

NMP acknowledges support from the European Research Council, ERC Ideas Starting grant n. 279985 "BIHSNAM: Bio-inspired Hierarchical Super Nanomaterials" and ERC Proof of Concept grant n. 619448 "REPLICA2: Large-area replication of biological anti-adhesive nanosurfaces" and from the European Commission within the Graphene Flagship. FB acknowledges support from BIHSNAM. P.O. is supported by the FIRB "Futuro in Ricerca 2010" project (CUP code: D11J11000450001) funded by the Italian Ministry for Teaching, University and Research (MIUR) and by "A.Di.N-Tech." project (CUP code: D15E13000130003) funded by University of Torino and Compagnia di San Paolo in the framework of the "Progetti di ricerca di Ateneo 2012" scheme, which are gratefully acknowledged.